\documentclass[graybox]{svmult}

\usepackage{mathptmx}       
\usepackage{makeidx}         
\usepackage{graphicx}        
\usepackage{multicol}        
\usepackage[bottom]{footmisc}
\usepackage{amssymb}
\usepackage{color}

\makeindex             

\begin{document}

\title*{The Many-agent Limit of the Extreme Introvert- Extrovert Model }
\author{ Deepak Dhar, Kevin E. Bassler,  and R. K. P. Zia}
\institute{Deepak Dhar  \at Tata Institute of Fundamental Research, Homi Bhabha Road, Mumbai, India  \email{ddhar@theory.tifr.res.in}  \and Kevin E. Bassler \at Department of Physics, University of Houston, Houston, TX 77204 USA \email{bassler@uh.edu} \and R. K. P. Zia \at Department of Physics, Virginia Polytechnic Institute and University, Blacksburg VA 24061 USA \email{rkpz@vt.edu} }

%
%
\maketitle

\abstract*{We consider a toy model of interacting  introvert and extrovert agents introduced earlier by Liu et al [Europhys. Lett. {\bf 100} (2012) 66007].  The number of extroverts, and introverts is $N$ each. At each time step, we select an agent at random, and allow her to modify her state.  If an  extrovert is selected, she adds a link at random to an unconnected introvert. If an introvert is selected, she removes one of her links. The set of $N^2$ links evolves in time, and may be considered as a set of Ising spins on an $N \times N$  square-grid  with single-spin-flip dynamics.  This dynamics satisfies detailed balance condition, and the probability of different spin configurations  in the steady state can be determined exactly. The effective hamiltonian  has long-range multi-spin couplings that depend on the row and column sums of spins.   If the relative bias of choosing an extrovert over an introvert is varied, this system undergoes a phase transition from a state with very few links to one in which most links are occupied. We show that the  behavior of the system can be determined  exactly in the limit of large $N$.  The behavior of large fluctuations in the total numer of links  near the phase transition is determined.  We also discuss two variations, called egalitarian and elitist agents, when the agents preferentially add or delete links to their least/ most-connected neighbor. These shows interesting cooperative behavior.}

\abstract{We consider a toy model of interacting  extrovert and introvert agents introduced earlier by Liu et al [Europhys. Lett. {\bf 100} (2012) 66007].  The number of extroverts, and introverts is $N$ each. At each time step, we select an agent at random, and allow her to modify her state.  If an  extrovert is selected, she adds a link at random to an unconnected introvert. If an introvert is selected, she removes one of her links. The set of $N^2$ links evolves in time, and may be considered as a set of Ising spins on an $N \times N$  square-grid  with single-spin-flip dynamics.  This dynamics satisfies detailed balance condition, and the probability of different spin configurations  in the steady state can be determined exactly. The effective hamiltonian  has long-range multi-spin couplings that depend on the row and column sums of spins.   If the relative bias of choosing an extrovert over an introvert is varied, this system undergoes a phase transition from a state with very few links to one in which most links are occupied. We show that the  behavior of the system can be determined  exactly in the limit of large $N$.  The behavior of large fluctuations in the total numer of links  near the phase transition is determined.  We also discuss two variations, called egalitarian and elitist agents, when the agents preferentially add or delete links to their least/ most-connected neighbor. These shows interesting cooperative behavior.}

\section{Introduction}
\label{sec:1}
In recent years, there has been a lot of interest in the study of networks.  many different types of networks have been studied: transportation networks like the railways or airlines networks \cite{railways}, chemical reaction networks in a cell \cite{cell}, power grids etc. \cite{powergrid}. A good review of the general theory may be found in \cite{dorogovetsev, scalefree}.  In studies of networks in social sciences, some examples are the citation network of scientific publications\cite{citations}, small-world networks \cite{smallworld}. An important question that has attracted a lot of attention is the time evolution of social networks, in which the number of friends a particular agent interacts with evolves in time.  In this context, a very interesting toy model was introduced by Liu et al \cite{LSZ12}, called the extreme introvert extrovert model.  Numerical studies of this model revealed a phase transition, where the fractional number of links present undergoes a jump from a value near $0$ to a value near $1$, as a parameter in the model is varied continuously \cite{LSZ13,LSZ14,BLSZ15}. Recently, we have shown that this model can be solved exactly in the limit of large number of agents \cite{BDZ15}. This article is a brief  pedagogical account of these results.  For details, the reader is referred to the original publication. 

\section{Definition of the Model}
\label{sec:2}

We consider a group consisting of $N$ introvert and $N$ extrovert agents. The agents can add or delete links  connecting them to other agents. In so doing, in our model,  they do not have to get permission from the agent  to whom the link is being added or from whom the link is being removed.  The introverts are assumed to be comfortable with only a few links, and extroverts like many.  In  a general model, the nember of links an introvert likes to have is $k_I$, and an extrovert likes $k_E$, with $k_I < k_E$. We will consider the extreme case where $k_I=0$, and $k_E=\infty$.  Thus, and introvert does not like any links, and will delete a link, given an opportunity.  On the other hand, an extrovert will try to add a link whenever possible.  This model has bee called  the eXtreme Introvert-Extrovert model(XIE) \cite{LSZ12,LSZ13,LSZ14}.

A configuration of the system at any time is specified completely by an $N \times N$ matrix ${\mathbb A}$ whose entries are $A_{ij}$, with $A_{ij} =1$, if there is link between the $i$-th introvert, and the $j$-th extrovert, and $A_{ij} = 0$ otherwise.  The total number of configurations is $2^{N^2}$.
The model undergoes a discrete-time Markovian evolution defined by the following rules: At each time step, select an agent at random, and allow her to change the number of links connecting her to other agents. Any introvert has a probability $\frac{1}{(1+z)N}$, of being selected, and an extrovert has a probability $\frac{z}{(1+z)N}$ of being selected. Then, $z < 1$ corresponds to a bias favoring  introverts, and $z > 1$  favors extroverts.  If an introvert is selected, and has at least one link to an extrovert, she deletes one of the links connecting her to other agents at random. If she has no links, she does nothing, and the configuration remains unchanged. If an extrovert is selected, she will add a link to one of introverts not already linked to her.  If she already has all links present, she does nothing.

\section{Steady State of the XIE model}
We may think of the $N^2$ binary variables $A_{ij}$ as Ising variables placed on an $N \times N$ square grid, and then the update rule corresponds to single-spin-flip dynamics of the model.  Note that our rules have the symmetry of changing introverts to extroverts, and $A_{ij} \leftrightarrow 1 - A_{ij}, z \leftrightarrow 1/z$.  This corresponds to the Ising model having spin-reversal symmetry with $z \leftrightarrow 1/z$. 

In general, given some such set of update rules, it is very difficult to determine the probabilities of different configurations in the steady state exactly. The remarkable result about the XIE model is that in this case, the probabilities of transition between configurations ${\mathcal C}$ and ${\mathcal C}'$ satisfy the detailed balance condition, and one can write down the probability of different configurations in the steady state exactly. For the configuration ${\mathcal C}$, in which the $i$-th introvert has degree $p_i$, and the $j$-th extrovert has degree $q_j$, the steady-state probability ${\rm Prob}^*({\mathcal C})$ has  a very pleasing form \cite{LSZ12}

\begin{equation}
\mathcal{P}^{\ast }( {\mathcal C}) =\frac{1}{\Omega(z) }%
z^{\sum_i p_i} \prod\limits_{i=1}^{N}\left( p_{i}!\right)
\prod\limits_{j=1}^{N}\left( N -q_{j}\right)! \label{P*}
\end{equation}
where $\Omega(z)$ is a normalization constant.

We may define the negative logarithm of this probability as the `energy' of the configuration ${\mathcal C}$, giving
\begin{equation}
H_{eff}({\mathcal C}) = -\sum_{i=1}^N \log p_i! -\sum_{j=1}^N \log (N -q_j)! -\log(z) \sum_i p_i
\end{equation}

We see that the effective hamiltonian has long-range couplings, and the energy of a configuration depends only on the row- and column- sums of the square array ${\mathbb A}$. Also, the energy function is non-extensive: the energy of the configuration with all links absent varies as $ - N^2 \log N$.  This non-extensivity causes no real problems, as all probabilities are well-defined, and averages of observables in steady state are well-behaved.

Monte Carlo simulations of the XIE model have shown that,  for large $N$, the system seems to undergo a  transition from a few-links phase for $z <1$ to a phase in which almost all links present for $z >1$. In the few-links phase, the average number of links per agent remains finite, of order 1, even as $N$ tends to infinity, with fixed $z < 1$. Conversely, in the link-rich phase for $z >1$, the average number of links per agent is nearly $N$, and the difference of this number from $N$ remains finite, as $N$ is increased to infinity.

The fact that energy depends only on the row- and column- sums, and thus only on $2N$ variables, instead of the $N^2$ variable $A_{ij}$ explicitly, suggests that some kind of mean-field treatment may be exact for this problem. This turns out to be true, as we proceed to show, but the treatment needs some care, as the number of variables, and hence also their conjugate fields,  tends to infinity, in the thermodynamic limit. 
If the energy of the system depended only on the total number of links in the system, a single mean-field variable conjugate to the net magnetization would have been sufficient.

\section{Asymptotically exact perturbation theory}

We consider the low-density phase ($ z< 1$) first. The case $z >1$  is equivalent to this by the Ising symmetry discussed above.
In the low density phase, the typical  degree  $q_j$ of the $j$-th extrovert is much less than  $ N$. Then we have $ (N-q_j)! \approx N! N^{-q_j} $. This suggests that we write  for all $q\geq 0$
\begin{equation}
(N-q)! / N! =  N^{-q} F(q,N)
\end{equation}
with 

\begin{equation}
F(q,N) = \prod_{r=1}^{q} \left( 1 -\frac{r-1}{N} \right)
\end{equation}
For $q \ll N$, $ F(q,N) $ is nearly equal to $1$.  Then, since $\sum_j q_j = \sum_i p_i$, we can 
write the effective Hamiltonian for the random XIE model as 
\begin{equation}
{\mathcal H}_{eff} ={\mathcal H}_{0}+{\mathcal H}_{int}
\end{equation}%
where 
\begin{equation}
{\mathcal H}_{0}=-\sum_{i}\left[ \ln \left( p_{i}!\right) +p_{i} \ln 
{\frac{z}{N}} \right] -N \ln (~N!)  \label{H0}
\end{equation}
and
\begin{equation}
{\mathcal H}_{int}=-\sum_{j}\ln F(q_{j},N)  \label{Hint}
\end{equation}

If we ignore the effect of the the "perturbation term " ${\mathcal H}_{eff}$, different introverts are independent, and one can sum over states of each introvert separately. This gives
\begin{equation}
\nonumber
\Omega _{0}=(N!)^{N}[\omega _{0}]^{N}
\end{equation}%
with 
\begin{equation}
\nonumber
\omega_{0}=\sum_{k} z^{k} F(k,N_{E})  \label{omega0}
\end{equation}%

For large $N$, $F$ tends to $1$, and we get
\begin{equation}
\omega_0 = 1 + z + z^2 + z^3 + ... = \frac{1}{1-z},
\end{equation}
which gives
\begin{equation}
 \log \Omega(z) = N ~\log (N!) + ~N ~\log ~(\frac{1}{1-z}) + \dots .
\end{equation}

In a systematic perturbation theory, we need to determine the behavior of the steady state of the  system under ${\mathcal H}_0$. This is easily done. In particular, we can determine the degree distribution of introverts and extroverts. It is easily seen that for large $N$, the probability that an introvert has degree $r$ has an exponential distribution : 
${\rm Prob~(introvert ~ has ~degree~} r) = ( 1 -z) z^r$.   Here, the $p_i!$ factor in the weight of a configuration makes the usually expected Poisson form into  a simple exponential. However,  the degree distribution of the $j$-th  extrovert is a  sum of $N$ mutually independent variables $A_{ij}$, hence it remains  a  Poisson distribution.  Clearly the  mean degree of extroverts  is same as the mean degree of introverts, so the Poisson dstribution has  mean $\frac{z}{(1-z)}$, which determines it completely.\\

To lowest order in $(1/N)$, $\log F(q,N) = -q(q-1)/(2N)$.  Thus, while the interaction hamilonian has different values for different configurations, and thus not a trivial c-number term, it is a sum of $N$ different weakly correlated terms, and its mean is ${\mathcal O}(1)$, and fluctuations about the mean are smaller. It is easily seen that they are  ${\mathcal O}(N^{-1/2})$,
giving

\begin{equation}
\log \Omega(z) = N ~\log (N!) + ~N ~\log ~(\frac{1}{1-z}) +  {\mathcal O}(1), {\rm ~for ~} z < 1.
\end{equation}

For $z > 1$, similar analysis, or the introvert-extrovert flip symmtery can be used to deduce that
\begin{equation}
\log \Omega(z) = N ~ \log (N!) + N^2 \log z + N ~ \log  ~(\frac{1}{1-1/z}) + {\mathcal O}(1), {\rm ~for ~} z > 1.
\end{equation}

This is a remarkable result. Clearly, we get asymptotically exact result for $\log \Omega(z)$  up to the linear order in $N$ using the hamiltonian ${\mathcal H}_0$. The effect of ${\mathcal H}_{int}$ is only a term of ${\mathcal O}(1) $ in $\Omega(z)$. In particular,  in the large-$N$ limit, the density of links is $0$ for $z < 1$, and $1$ for $z >1$.

We note that these results are consistent with a scaling ansatz 

\begin{equation}
\frac{ \left[ \Omega(z)\right]^{1/N}}{ N!}  = N^a f(\epsilon N^b)
\end{equation}
where $  z = \exp( -\epsilon), a = ~b = 1/2$, and $f(0) = $ a finite constant, and $f(x)$ is continuous at $x =0$. For large positive $x$, $f(x) \sim 1/x$. For $ z > 1$, a similar form is obtained by the inversion symmetry, but the scaling functions for $\epsilon > 0$ and $\epsilon < 0$ are different, reflecting the `first-order nature' of the transition.

\section{Variants of XIE with preferential attachment}

It is interesting to consider some variations on this general theme. We consider the case when the agent 
does not choose which link to add (or delete) at random, as done in the XIE model in section 2, but decides on the basis of 
knowledge of degree of the receiving node. We consider two variations.

\textit{Egalitarian agents:}  Here  extrovert  agents realize that
the introverts regard links as burden, and attempts to distribute this
burden as evenly as possible, and would add a link to the least connected introvert. In case
Similarly, an introvert would cut a link to the \textit{most} connected
extrovert, as this action would make the other extroverts more equal. 

\textit{Elitist agents:}  Here, we consider the
opposite extreme. In this case, an extrovert prefers the most `sociable'
introvert, and adds a link to the \textit{most} connected of the available
introverts. Similarly, an introvert cuts a
link to the \textit{least} connected available extrovert.

These variations have a strong effect on the degree distribution in the steady state.
Let us discuss egalitarians first. Then, at any time, the degree distribution of an agent will
have only two possible values: $k$ or $(k+1)$, for some possibly time-dependent $k$. 
In the low density regime of this egalitarian XIE model, there are only a small
number of contacts. It is easy to see that in the large-$N$ limit,  in the steady state, we have  $k =0$, and 
fractional number of introverts with exactly $1$  contact is $z$. For the degree distribution of extroverts, it is easy to see that degree distribution is Poisson, with a mean value that increases with $z$.  For $z>1$, the only possible values of degree of an introvert are $N-1$ and $N$. 

\begin{figure}[tbp]
\centering
\includegraphics[width=3.5in]{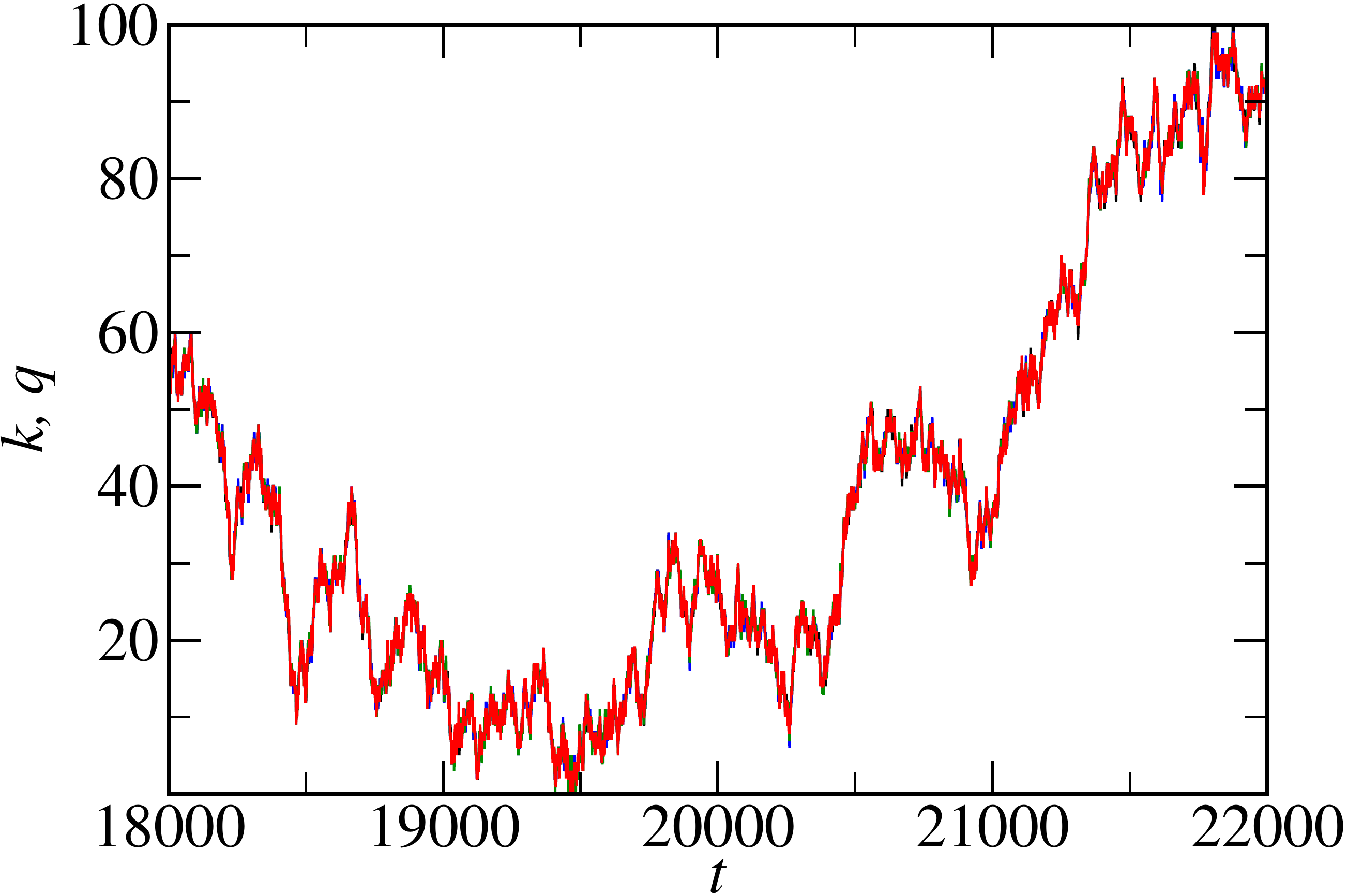}
\caption{Time trace of the degrees of two introvert  $k$
(black, blue), and of two extroverts, $q$ (red, green) in a critical egalitarian XIE with $N=100$.  The graphs fall on top of each other, and are not distinguishable.  Here the unit of $t$ is a sweep. Taken from \cite{BDZ15}.}
\label{4maxagents}
\end{figure}

The behavior of the degree distribution at the phase transition point $z=1$ is particularly interesting.  Here the fractional number of links can vary from $0$ to $1$, and the degree of an agent vary from $0$ to $N$. However, the agents, by their cooperative behavior ensure that the inequality in the soceity always remains low.  This is shown in Fig. \ref{4maxagents} , where we plot the time-dependent degree of two introverts, and two extroverts.  While the actual degree varies in an unpredictable way, the four curves fall on top of each other, and are not distinguishable in the plot.

\ 

We now discuss the elitists case.  To recall, here  the extrovert agents prefer to link to  one most sociable of the introverts ( most connected), and  introverts, to keep inversion symmetry,  are  assumed to delete their link to  the least connected extrovert.
This generates an interesting instability:
Say, start with all links absent. Then, an extrovert will add a link to some randomly chosen introvert. Then all later extroverts selected at subsequent times  will choose to link to the same agent.  Her degree will tend to become $N$. Then the extroverts will choose a new "star' to link to.  The degree distribution of introverts becomes very unequal. The question what  happens in the steady state?  

Interestingly, it turns out that in this case, for all $z$, the degree distribution is rather wide. Let us try to understand this. Clearly, argument sketched above needs elaboration. When a `star' introvert's degree becomes comparable to $N$, often an updating extrovert would be already connected to this introvert, and the time between changes of  degree  increases if the degree is closer to $N$.  Meanwhile, other new stars are already in the making. This produces a broad distribution of degrees.  If we inspect the incidence matrix ${\mathbb A}$ at different times during the evolution of the system, we do not see much pattern in it directly. 
However, if we look at the same matrix after permuting the rows and columns to matrix so that both introverts and extroverts are arranged  according to their degrees  in ascending order, we see much more structure. It is found, and easy to prove a posteriori, that in the sorted matrix, all $1$'s come in a single block with no holes, and similarly with zeroes [ Fig. 2].  There is a single staircase-shaped interface that separates 1's and zeros, and this fluctuates with time. Clearly, if we have this property at one time, it will be preserved by the dynamical rules.

The total number of accessible matrices in steady state is $ \leq {2N \choose N}  (N!)^2$, which is much less than the number of possible  $2^{N^2}$. Thus, the configurations are much simpler to describe, when the permutation symmetry between agents is factored out, and one works with the equivalence classes of configurations under the symmetry group. As time evolves, the interface describing the configuration will evolve (we have to reorder the agents according to their degree distribution at each time). One can write the evolution rules in terms of the interface model.  It can be seen that the interface will evolve by flipping at corners:  height can change at the maxima, or the minima, at  a rate proportional to length of side on the right [Fig. \ref{interface}]. But this model seems difficult to solve exactly.  On general grounds, since it is an interface fluctuation model where height can increase or decrease with equal probability, it may be expected to be in the universality class of Edwards-Wilkinson model, where the dynamical exponent is $2$, i.e. relaxation time for an interface of length $L$ varies as $L^2$. For the elitists case, this corresponds to relaxation time being approximately  ${\mathcal O}(1)$ sweeps (where one sweep of the system  is $L^2$ attempts which updates each link on the average once), for large $N$. 

For different values of the $z$, the slope of the interface changes, but qualitative behavior of relaxation remains the same. Hence, we find that the elitists self-organize into a critical state, where the degree distribution of agents is wide, for all $z$.

\begin{figure}[tbp]
\centering
\includegraphics[width=1.5in]{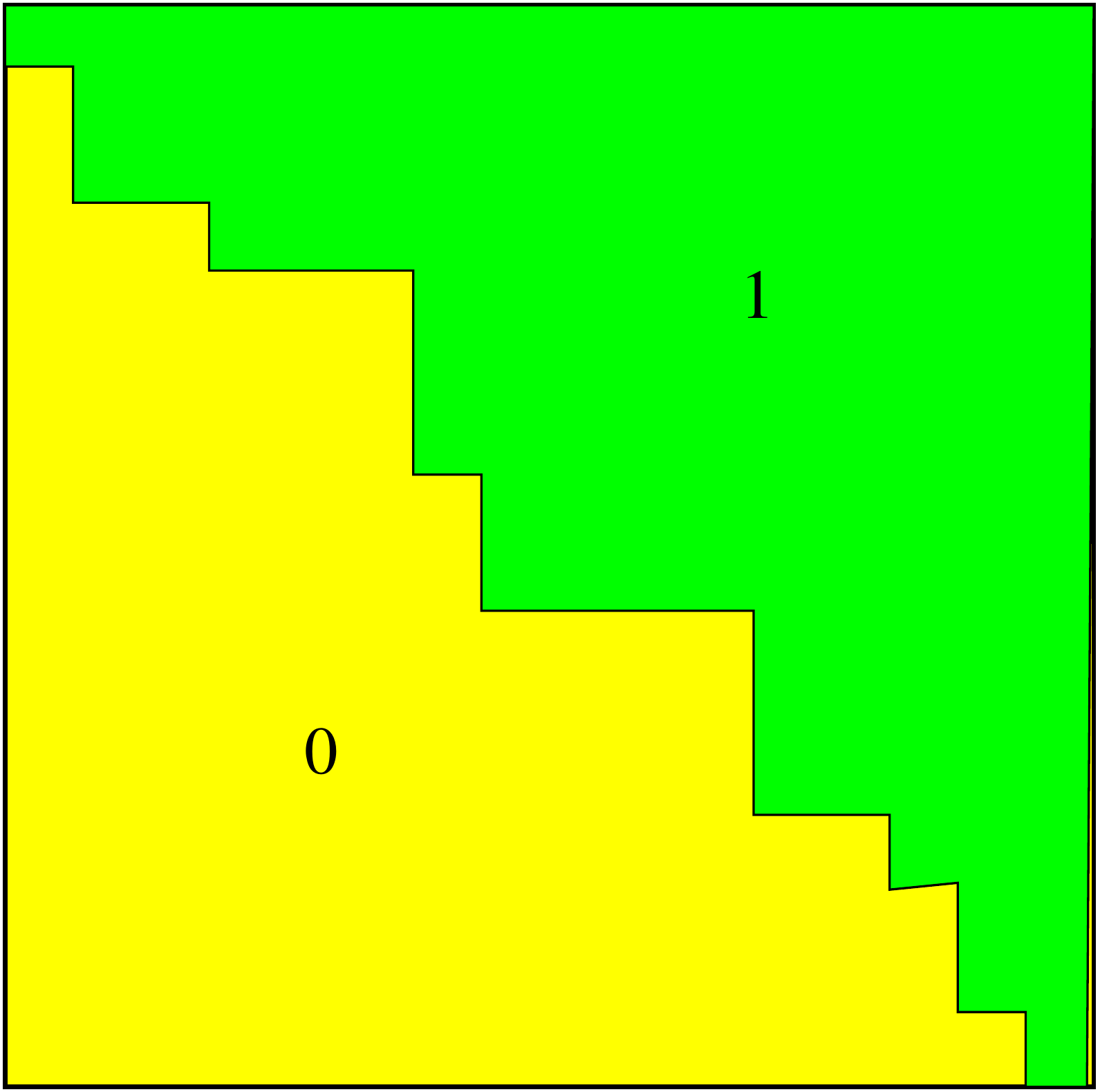}
\caption{The incidence matrix after reordering.  The 1's and zeroes are separated by a single staircase-like interface. Here green =1, yellow =0. } 
\label{min15050}
\end{figure}

~\\
\begin{figure}
\centering
\includegraphics[width=3.5in]{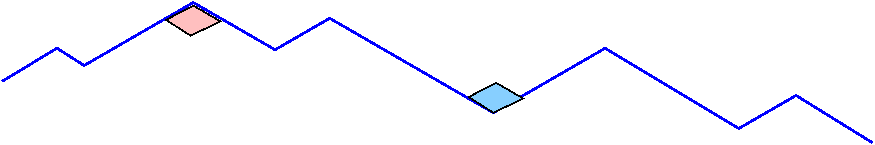}
\caption{Evolution rules for the interface: height can change at the maxima (pink diamond), or the minima(blue diamond), at  a rate proportional to length of side on the right.}
\label{interface}
\end{figure}

\section{Summary and Concluding Remarks}

In this article, we discuss a simple toy model of a dynamical networks, where the state of different links  keeps changing with time, but the system has  a non-trivial steady state. As a function of the bias parameter $z$, the system undergoes a phase transition from a state with few links to  a state with most links occupied. We showed that one can develop a perturbation theory,  which becomes asymptotically exact for large $N$, at first order of perturbation. The corresponding state has non-trivial many-body correlations. We also discussed variations of the basic XIE model where the agents attach to one of the  most- or least- connected  agents, and showed that the steady state can have a wide degree distribution of degrees of agents.  An  interesting open problem is a  theory  to calculate the scaling  function $f(x)$  in Eq.(14) exactly.  It is hoped that future research will lead to a better understanding of this model.

\begin{acknowledgement}
 DD and RKP  thank the Galileo Galilei Institute for Theoretical Physics
for hospitality and the INFN for partial support during the summer of 2014. This research is
supported in part by the Indian DST via grant DST/SR/S2/JCB-24/2005, and  the US NSF via grants DMR-1206839 and DMR-1507371. 

\end{acknowledgement}

\end{document}